\documentclass[twocolumn,prb,floats,amsmath,amssymb,superscriptaddress]{revtex4}
\usepackage{epsfig}

\usepackage{graphicx}
\usepackage{dcolumn}
\usepackage{bm}
\usepackage{chemarr}
\usepackage{multirow}
\usepackage{subfigure}
\usepackage{epstopdf}

\usepackage{color}
\definecolor{darkgreen}{rgb}{0,0.5,0}
\definecolor{blue}{rgb}{0,0,0.8}
\definecolor{lightblue}{rgb}{0.93,0.96,1}
\definecolor{darkblue}{rgb}{0.,0.,0.6}
\usepackage[colorlinks,linkcolor=blue,citecolor=blue,urlcolor=blue]{hyperref} 

\renewcommand\matrix[1]{\,\mathrm{\underline{#1}}\,}

\DeclareGraphicsExtensions{.png,.gif,.jpg,.pdf,.bmp}

\begin{document}

\title{Limitations of the stochastic quasi-steady-state approximation in open biochemical reaction networks\footnote{ Paper accepted by J. Chem. Phys. (Communication) }}

\author{Philipp Thomas}
\affiliation{Department of Physics, Humboldt University of Berlin, Newtonstr. 15, D-12489 Berlin, Germany}
\affiliation{School of Biological Sciences, University of Edinburgh, Edinburgh EH9 3JR, United Kingdom}

\author{Arthur V. Straube}
\affiliation{Department of Physics, Humboldt University of Berlin, Newtonstr. 15, D-12489 Berlin, Germany}

\author{Ramon Grima}
\affiliation{School of Biological Sciences, University of Edinburgh, Edinburgh EH9 3JR, United Kingdom}

\begin{abstract}
The application of the quasi-steady-state approximation to the Michaelis-Menten reaction embedded in large open chemical reaction networks is a popular model reduction technique in deterministic and stochastic simulations of biochemical reactions inside cells. It is frequently assumed that the predictions of the reduced master equations obtained using the stochastic quasi-steady-state approach are in very good agreement with the predictions of the full master equations, provided the conditions for the validity of the deterministic quasi-steady-state approximation are fulfilled. We here use the linear-noise approximation to show that this assumption is not generally justified for the Michaelis-Menten reaction with substrate input, the simplest example of an open embedded enzyme reaction. The reduced master equation approach is found to considerably overestimate the size of intrinsic noise at low copy numbers of molecules. A simple formula is obtained for the relative error between the predictions of the reduced and full master equations for the variance of the substrate concentration fluctuations. The maximum error is reached when modeling moderately or highly efficient enzymes, in which case the error is approximately $30\%$. The theoretical predictions are validated by stochastic simulations using experimental parameter values for enzymes involved in proteolysis, gluconeogenesis and fermentation.
\end{abstract}

\maketitle

It is well known that whenever transients in the concentration of a substrate species decay
over a much slower timescale than those of the enzyme species, one can invoke the quasi-steady-state approximation (QSSA) to considerably simplify the deterministic (macroscopic) rate equations.\cite{SegelSlemrod,SchnellMaini} The study by Rao and Arkin\cite{RaoArkin}  pioneered the use of the same approximation but on a mesoscopic level, i.e., applying a stochastic version of the approximation to obtain reduced chemical master equations. This approximation has since become ubiquitous in stochastic simulations of large biochemical reaction networks inside cells (see, for example, Refs.~\onlinecite{Gonze2008,Paulsson2000,Dupont2008,Guerriero2011}) although its range of validity is presently unknown. A plausible hypothesis is that the stochastic QSSA is valid in the same regions of parameter space where the deterministic QSSA is known to be valid. A handful of numerical studies \cite{Gonze2011,Sanft2011} have shown that for some choices of rate constants which are compatible with the deterministic QSSA, the differences between the reduced and full master equation approaches are practically negligible. However, none of these studies exclude the possibility that there exist regions of parameter space where the deterministic QSSA is valid but the stochastic QSSA exhibits large systematic errors in its predictions. In particular, one is interested in knowing how accurate are the predictions of the stochastic QSSA for the size of intrinsic noise, i.e., the size of fluctuations in concentrations, since such noise is known to play important functional roles in biochemical circuits.\cite{Eldar2010} Numerical approaches cannot easily answer such questions because the stochastic simulation algorithm, the standard method which exactly samples the trajectories of master equations,\cite{Gillespie1977} is computationally expensive.\cite{Gillespierev}

In this communication, we seek to develop a theoretical approach to answer the following question: Given that the rate constants are chosen such that the deterministic QSSA is valid, what are the differences between the predictions of the reduced and full master equations for the variance of the fluctuations about the mean concentrations? We obtain a formula estimating the size of these differences for the simplest biochemical circuit which embeds the Michaelis-Menten reaction and confirm its accuracy using stochastic simulations. We find, using physiological parameter values, that the reduced master equation approach can overestimate the variance of the fluctuations by as much as $\sim 30\%$.

We start by considering the Michaelis-Menten reaction with substrate input
\begin{align}
 \label{eqn:MMreaction}
 \xrightarrow{k_\text{in}} X_S, \quad X_S + X_E \xrightleftharpoons[k_{1}]{k_0} X_C \xrightarrow[]{k_2} X_E + X_P,
\end{align}
where  $X_i$ denotes chemical species $i$ and the $k$'s denote the associated macroscopic rate constants. The reaction can be described as follows. Substrate molecules (species $S$) are pumped into some compartment at a constant rate, they bind to free enzyme molecules (species $E$)  to form substrate-enzyme complexes (species $C$) which then either decay back to the original substrate and free enzyme molecules or else decay into free enzyme and product molecules (species $P$). The first reaction in (\ref{eqn:MMreaction}) could equally represent the production of substrate by a first-order chemical reaction provided the species transforming into substrate exists in concentrations large enough such that fluctuations in its concentration can be ignored. The sum of the concentrations of free enzyme and complex is a constant since the enzyme can only be in one of these two forms. Hence, all mathematical descriptions of the Michaelis-Menten reaction can be expressed in terms of just complex and substrate variables. On the macroscopic level, the QSSA proceeds by considering the case in which transients in the complex concentration decay much faster than those of the substrate. This condition of timescale separation is imposed by setting the time derivative of the macroscopic complex concentration to zero, solving for the steady-state complex concentration and substituting the latter into the rate equation for the substrate concentration which leads to the new rate equation
\begin{equation}
\frac{\partial}{\partial t} [X_S (t)] = k_{in} - \frac{k_2 [E_T] [X_S(t)]}{K_M + [X_S(t)]}, \label{eq2}
\end{equation}
where $[X_S(t)]$ is the substrate concentration at time $t$, $K_M = (k_1 + k_2)/k_0$ is the Michaelis-Menten constant, and $[E_T]$ is the total enzyme concentration, i.e., the sum of the concentration of free enzyme, $[X_E(t)]$, and of the concentration of complex, $[X_C(t)]$, which is a constant as previously mentioned. Note that the notation $[X_i]$ (without explicit dependence on $t$) will be reserved for the steady-state concentration of species $i$. We assume that at $t = 0$ we are in steady-state conditions; note that the system (\ref{eqn:MMreaction}) is guaranteed to have a stable steady-state if the condition $k_{in}/k_2 [E_T] < 1$ is satisfied. Linear stability analysis of the full rate equations describing (\ref{eqn:MMreaction}) shows that the timescale for the decay of transients in the substrate concentrations is $\tau_s = (k_0 [X_E])^{-1}$, while the timescale for the decay of transients in the complex concentrations is $\tau_c = (k_0 [X_S] + k_1+k_2)^{-1}$. Hence, the criterion for the validity of the QSSA on the macroscopic rate equations (the deterministic QSSA), i.e., for the validity of Eq.~(\ref{eq2}), reads $\tau_s / \tau_c = \gamma =  ([X_S] + K_M)/[X_E] \gg 1$ (see also Ref.~\onlinecite{Stoleriu}). This condition implies that timescale separation is possible over the whole range of macroscopic substrate concentrations whenever the total enzyme concentration is much smaller than the Michaelis-Menten constant.

The question that we address in the rest of this communication is the following: Given that the condition $\gamma \gg 1$ is satisfied, what is the variance of the noise about the mean concentrations as predicted by the reduced and full master equations?

The stochastic QSSA method implicitly starts by deducing that Eq.~(\ref{eq2}) is effectively the rate equation one would associate with a system of two chemical processes
\begin{align}
\xrightarrow{k_\text{in}} X_S, \quad X_S \xrightarrow{k'} X_P, \label{eq3}
\end{align}
where $k'$ is an effective (time-dependent) rate constant equal to $k_2 [E_T] / (K_M + [X_S(t)])$. Note that while the first reaction is elementary, the second is clearly not, since it can clearly be broken down into a set of more fundamental constituent reactions. Given the reduced set of reactions (\ref{eq3}) one can then construct a reduced master equation for the set of reactions (\ref{eqn:MMreaction}) (see Ref.~\onlinecite{vanKampen} and supplementary material for the construction of master equations)
\begin{align}
\frac{\partial}{\partial t} P(n_S,t) = &\Omega k_{in} (E_S^{-1}-1) P(n_S,t) \nonumber  \\ &+(E_S^{+1}-1) \frac{k_2 [E_T] n_S}{K_M + n_S / \Omega} P(n_S,t), \label{eq4}
\end{align}
where $\Omega$ is the compartment volume in which the reactions are occurring, $n_S$ is the absolute number of substrate molecules, $P(n_S,t)$ is the probability that the system has $n_S$ substrate molecules at time $t$ and $E_S^{m}$ is the step operator which upon acting on a function of $n_S$ changes it into a function of $n_S + m$ (Ref.~\onlinecite{vanKampen}). We note and emphasize that the physical basis of this master equation is not clear because such equations have been derived from first principles for elementary reactions,\cite{Gillespie1992,Gillespie2009} while (\ref{eq3}) involves a non-elementary reaction. Equation~(\ref{eq4}) is simply written by analogy to what one would write down for (\ref{eq3}) if both reactions were elementary and hence its legitimacy is \textit{a priori} doubtful.

Now we want to use this master equation to deduce the variance of the noise in the macroscopic substrate concentrations. It is well known that in the macroscopic limit, the master equation for monostable chemical systems can be approximated by a linear Langevin equation, an approximation called the linear noise approximation (LNA).\cite{vanKampen,ElfEhrenberg} For systems with absorbing states or exhibiting multimodality, the LNA will not usually give accurate results (see, for example, Ref.~\onlinecite{LNAbreak}) but its application to our example, the Michaelis-Menten reaction with substrate input, is not problematic since this reaction is only capable of monostable behavior. The steps to construct the LNA for a general monostable chemical reaction system are summarized in the supplementary material. Here, we will simply state the results of this recipe when applied to the master equation, Eq.~(\ref{eq4}). The Langevin equation approximating Eq.~(\ref{eq4}) in the macroscopic limit of large molecule numbers and in steady-state conditions is
\begin{equation}
\frac{\partial}{\partial t} \eta_S(t) = -\frac{k_2}{\gamma} \eta_S(t) + \sqrt{\frac{2 k_2 [X_S]}{\Omega \gamma}\biggl(1+\frac{[X_S]}{K_M} \biggr)}\Gamma(t), \label{eq5}
\end{equation}
where $\eta_S(t)$ denotes the fluctuations about the macroscopic steady-state substrate concentration defined as $\eta_S(t)= n_S(t) / \Omega - [X_S]$ and $\Gamma(t)$ is white Gaussian noise defined by $\langle \Gamma (t) \rangle = 0$ and $\langle \Gamma (t)\Gamma (t') \rangle = \delta(t - t')$. From Eq.~(\ref{eq5}), one can show (see supplementary material) that the variance of concentration fluctuations $\eta_S(t)$ in steady-state conditions is given by
\begin{align}
\sigma_{sLNA}& = \frac{[X_S]}{\Omega}\left(1+\frac{[X_S]}{K_M} \right), \label{eq-sigmasLNA}
\end{align}
where the subscript ``sLNA'' stands for ``LNA of the master equation reduced using the stochastic QSSA''. Note that $[X_S]$ in Eqs.~(\ref{eq5}) and (\ref{eq-sigmasLNA}) is the steady-state substrate concentration obtained by solving for $[X_S]$ from Eq.~(\ref{eq2}) with time derivative set equal to zero.

We shall now derive expressions for the variance of substrate concentration fluctuations using the full master equation approach. The steps of this method are as follows: (i) one writes down the master equation for the elementary chemical processes (\ref{eqn:MMreaction}), (ii) the two Langevin equations for the complex and substrate fluctuations are obtained using the LNA of the master equation, (iii) expressions are found for the variance of complex and substrate concentration fluctuations in steady-state conditions, (iv) the limit $\gamma \gg 1$ is taken of the expressions derived in step (iii), leading to the final expressions for the variance of substrate fluctuations about the steady-state substrate concentration solution of the rate equation, Eq.~(\ref{eq2}). We note that this method, unlike the first one, does not make any assumptions about the validity of an ad-hoc reduced master equation since it is based on the master equation for elementary processes and hence is guaranteed to be correct.\cite{com} We now proceed to put this systematic recipe in practice.

The master equation for the four elementary chemical processes given by (\ref{eqn:MMreaction}) is
\begin{align}
\partial_t  P(n_S,n_C,t) &=\biggl[\frac{k_0}{\Omega} (E_S^{+1}E_C^{-1}-1) n_S (n_T - n_C)  \nonumber \\& + \Omega k_{in} (E_S^{-1}-1) + k_1 (E_S^{-1}E_C^{+1}-1) n_C  \nonumber \\ &+ k_2 (E_C^{+1}-1) n_C\biggr]  P(n_S,n_C,t), \label{eq7}
\end{align}
where $n_C$ is the absolute number of complex molecules and $n_T$ is the absolute total number of molecules of enzyme in free and complex form. Note that $n_T$ is a constant equal to $[E_T] \Omega$. In the macroscopic limit, the master equation, Eq. (\ref{eq7}), can be approximated by a pair of Langevin equations as given by the LNA (see supplementary information)

{\footnotesize
\begin{align}
\frac{\partial}{\partial t} & \begin{pmatrix}
\eta_{C}(t) \\
\eta_{S}(t)
\end{pmatrix}=k_0 \begin{pmatrix}
 -(K_M + [X_S]) &  [X_E] \\
 K_1 + [X_S] & - [X_E]
\end{pmatrix} \begin{pmatrix}
 \eta_{C}(t) \\
\eta_{S}(t)
\end{pmatrix} \nonumber \\
& + \sqrt{\Omega} \begin{pmatrix}
0 & \sqrt{k_0 [X_E] [X_S]} & -\sqrt{k_1 [X_C]} & -\sqrt{k_2 [X_C]} \\
 \sqrt{k_{in}} & -\sqrt{k_0 [X_E] [X_S]} & \sqrt{k_1 [X_C]} & 0
\end{pmatrix} \vec{\Gamma}(t), \label{eq8}
\end{align}}

\noindent where $K_1 = k_1/k_0$, $\eta_C(t)$, and $\eta_S(t)$, respectively, denote the fluctuations about the macroscopic steady-state complex and substrate concentrations and $\vec{\Gamma}(t)$ is a $4$ dimensional vector whose entries are white Gaussian noise with the properties $\langle \Gamma_i(t) \rangle = 0$ and $\langle \Gamma_i (t)\Gamma_j (t') \rangle = \delta_{i,j} \delta(t - t')$. Using Eq.~(\ref{eq8}), one can show (see supplementary information) that the variance of substrate concentration fluctuations $\eta_S(t)$ in steady-state conditions is given by
\begin{align}
\sigma_{LNA}& = \frac{[X_S]}{\Omega}\left(1+\frac{[X_S]}{K_M} \frac{K_1 + [X_S]}{K_M + [X_S]} \frac{\gamma}{1+\gamma} \right) \nonumber \\ & \xrightarrow{\gamma \gg 1}  \frac{[X_S]}{\Omega}\left(1+\frac{[X_S]}{K_M} \frac{K_1 + [X_S]}{K_M + [X_S]} \right), \label{eq-sigmaLNA}
\end{align}
where in the last step we took the limit of $\gamma \gg 1$, corresponding to the condition in which the deterministic QSSA Eq.~(\ref{eq2}) is valid.

Comparing Eqs.~(\ref{eq-sigmasLNA}) and (\ref{eq-sigmaLNA}), we see that the two are not generally equal to each other except in the case $\beta = k_2 / k_1 \ll 1$. From a fundamental point of view, this disagreement implies that the reduced master equation does not obey the generalized fluctuation-dissipation theorem of nonequilibrium physics \cite{vanKampen,Keizer} and that hence it is flawed. More importantly, we observe that the condition $\beta = k_2 / k_1 \ll 1$ is not equivalent to the quasi-steady-state condition $\gamma \gg 1$. The former condition is consistent with the enzyme-substrate complex being in thermodynamic equilibrium with free enzyme and substrate, a condition which is difficult to uphold in open systems since they are characterized by nonequilibrium steady states. While the quasi-steady-state condition can easily be satisfied in open systems since it is only required that the total enzyme concentration is much less than the Michaelis-Menten constant. Hence, we can conclude that for open systems, the stochastic QSSA based on Eqs.~(\ref{eq3}) and (\ref{eq4}) is {\em not} the legitimate stochastic equivalent of the deterministic QSSA.

There are two possible hypothetical scenarios which would imply that the stochastic QSSA is perhaps still a very good general method to estimate the size of the concentration fluctuations. The first case would be if experimental evidence showed that for many enzymes it just happens that $\beta \ll 1$. The second case would be if experimental evidence showed no such restriction on $\beta$ but nevertheless the difference between the variance prediction of the reduced and full master equations is so small as to be negligible. We now consider each case.

A perusal of the experimental data available in the literature shows that there are very few studies which simultaneously report values of $k_1$ and $k_2$, the data required to estimate $\beta$. The vast majority of studies report values for $K_M$, a considerable number report $k_2$ and a small percentage report both $k_2$ and $K_M$.\cite{Bar-Even2011} Now the ratio $k_2/K_M$, frequently called the enzyme efficiency,\cite{Fersht} is defined as
\begin{equation}
\Theta = \frac{k_2}{K_M} = \frac{\beta}{1 + \beta} k_0. \label{Theta}
\end{equation}
The recent study by Bar-Even \textit{et al.} \cite{Bar-Even2011} based on mining the Brenda \cite{Pharkya2003} and KEGG databases \cite{Kanehisa2008} concluded that for most enzymes $\Theta$ lies in the range $10^3 - 10^6 \,M^{-1} s^{-1}$. It is also known that the association constant $k_0$ takes values in the range\cite{Fersht} $10^6 - 10^9 \, M^{-1} s^{-1}$. We can conclude from these two pieces of data and using Eq.~(\ref{Theta}) that the range of $\beta$ for most enzymes is between $0$ and some number which is much greater than $1$ and that hence on the basis of experimental data one cannot argue for the general validity of the stochastic QSSA.

\begin{figure*}[!hbt]
\includegraphics[width=0.7\textwidth]{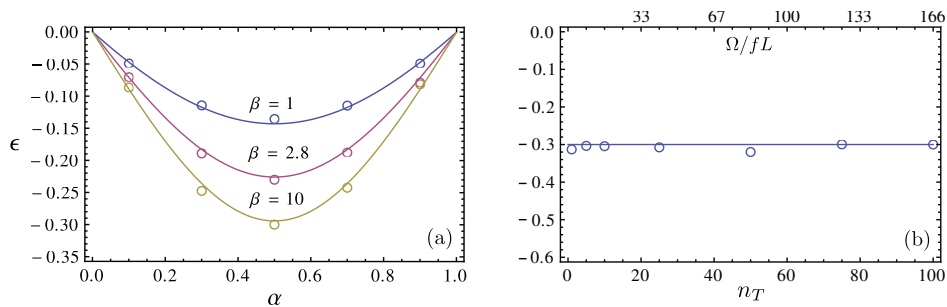}
\caption{(a) Plot of the fractional relative error, $\epsilon$, in the variance of fluctuations as predicted by the reduced master equation obtained from the stochastic QSSA versus the enzyme saturation parameter $\alpha$. The solid lines show the theoretical prediction Eq.~(\ref{eq11}) for three different values of $\beta$: $1$ (blue), $2.8$ (magenta) and $10$ (yellow-green). These three cases are respectively consistent with the enzymes being Chymotrypsin or Malate dehydrogenase, Lactate dehydrogenase and a highly efficient enzyme such as $\beta$-Lactamase. The data points show stochastic simulations using the Gillespie algorithm for reactions involving a total of $100$ enzyme molecules and total enzyme concentrations in the nano and millimolar range (see supplementary information for details). (b) Plot of the fractional relative error $\epsilon$ versus the total number of enzyme molecules $n_T$ for the case $\alpha = 0.5$ and $\beta = 10$. The data points are obtained from stochastic simulations. The solid line is simply a guide to the eye. The total number of enzymes is varied at constant total enzyme concentration. The ratio of substrate and complex decay timescales, $\gamma$, is greater than 10 in all cases shown in (a) and (b) which enforces the validity of the deterministic QSSA. Details are in supplementary information.}
\label{fig12}
\end{figure*}

Of course, as previously mentioned, it could still happen that even though there is no restriction on $\beta$, that the variance as predicted by the stochastic QSSA and the true variance are negligibly small. We can test this hypothesis quantitatively by using Eqs.~(\ref{eq-sigmasLNA}) and (\ref{eq-sigmaLNA}) to derive the fractional relative error $\epsilon$ in the variance prediction of the stochastic QSSA
\begin{equation}
\epsilon = \frac{\sigma_{LNA}-\sigma_{sLNA}}{\sigma_{LNA}} = \frac{-(1-\alpha)\alpha \beta}{1 + \beta (1 - \alpha(1-\alpha))}, \label{eq11}
\end{equation}
where $\alpha = k_{in}/(k_2 [E_T])$, a non-dimensional quantity which can take values between $0$ and $1$ as previously mentioned in the discussion after Eq.~(\ref{eq2}). Furthermore, it can be shown using Eq.~(\ref{eq2}) that at steady-state one has $[X_E] = [E_T] (1 - \alpha)$, and $[X_C] = [E_T] \alpha$, from which we can deduce that $\alpha$ is a measure of how saturated is the enzyme with substrate. Note that Eq.~(\ref{eq11}) shows that the relative error tends to zero as $\alpha \rightarrow 0$ and $\alpha \rightarrow 1$ and that hence the reduced master equation provides a correct prediction of the size of the substrate fluctuations whenever the free enzyme or complex concentrations are very small (similar results have been obtained by Mastny {\it et al.}\cite{Mastny2007} for the Michaelis-Menten reaction with no substrate input; however their results are not for general $\alpha$ and $\beta$ and do not enforce the validity of the deterministic QSSA; see later for discussion). In Fig.~\ref{fig12}(a), the solid lines illustrate the predictions of Eq.~(\ref{eq11}) for three different values of $\beta$: (i) $1$, (ii) $2.8$, and (iii) $10$. Case (i) utilizes experimental data for the enzymes Chymotrypsin and Malate dehydrogenase with respective substrates Acetyl-L-tryptophan and NADH,\cite{Lodola1978,Renard1973} while case (ii) is based on data for the enzyme Lactate dehydrogenase with substrate pyruvate.\cite{Boland1975} These enzymes are respectively involved in proteolysis, gluconeogenesis and the conversion of pyruvate (the final product of glycolysis) to lactate in anaerobic conditions. Case (iii) showcases the largest possible error made by the stochastic QSSA; this is consistent with a highly efficient enzyme such as $\beta$-Lactamase for which $\Theta$ is of the same order of magnitude as the maximum possible association rate constant\cite{Fersht} $k_0 \sim 10^8-10^9 s^{-1} M^{-1}$. The theoretical predictions of our LNA based method are confirmed by stochastic simulations of the master equations, Eq.~(\ref{eq4}) and Eq.~(\ref{eq7}), using Gillespie's algorithm \cite{Gillespie1977} [data points in Fig.~\ref{fig12}(a)]. Note that the maximum possible percentage error is about $30\%$, which is significant. Also note that the maximum error in all cases is reached at $\alpha = 1/2$, namely, when the enzyme is half saturated with substrate which occurs when the substrate concentrations are equal to the Michaelis-Menten constant $K_M$ (this is the case for most enzymes of the glycolytic pathway \cite{inhib}); for substrate concentrations much smaller or larger than $K_M$, the error is negligible.

The LNA is, strictly speaking, valid for large volumes or, equivalently, in the limit of large number of molecules \cite{vanKampen,Grima2010} and hence one could argue that our theoretical formula Eq.~(\ref{eq11}) is of limited validity inside cells, where molecule numbers can be quite small.\cite{GrimaSchnell} Figure~\ref{fig12}(b) shows the results of stochastic simulations for the case $\alpha = 0.5$ and $\beta = 10$ using a total number of enzyme molecules $n_T$ varying between $1$ and $100$ molecules. Note that the error $\epsilon$ is practically constant at $30 \%$, the value predicted by the LNA and shown in Fig.~\ref{fig12}(a). This suggests that the estimates provided by our method are accurate even for low copy number conditions.

Our study has focused on the most common type of stochastic QSSA in the literature which is heuristic in nature and hence the question regarding its validity. There are a class of alternative model reduction techniques \cite{Mastny2007,Srivastava2011} based on singular-perturbation analysis (sQSPA and sQSPA-$\Omega$) which are rigorous and whose validity is not under question. For the Michaelis-Menten reaction without substrate input, these methods lead to a reduced master equation of the same form as the heuristic stochastic QSSA whenever the free enzyme or complex concentrations are very small (see Table II of Ref.~\onlinecite{Mastny2007}). This implies that for such conditions the error in the predictions of the stochastic QSSA should be zero, a result which is also reproduced by our method. However, note that though these concentration conditions can be compatible with the deterministic QSSA they are not synonymous with it. The sQSPA methods do not lead to a reduced master equation for parameters consistent with the deterministic QSSA and hence cannot make statements regarding the accuracy of the heuristic stochastic QSSA in such conditions. Our contribution fills this important gap by deriving an explicit formula for the error in the predictions of the stochastic QSSA, i.e., Eq.~(\ref{eq11}), for all parameters values consistent with the deterministic QSSA. We finish by noting that a recent study by Gonze \textit{et al.} \cite{Gonze2011} also studied the reaction system (\ref{eqn:MMreaction}) using numerical simulations and found little difference between the predictions of the stochastic QSSA and the full master equation. The study used values of $\beta = 0.1$ (see Table $7.2$ in Ref.~\onlinecite{Gonze2011}) and hence in the light of our results, it is clear why they observed high accuracy of the stochastic QSSA. However, as we have shown, this is not the general case: many enzymes have large $\beta$ and hence discrepancies of the order of few tens of percent between the predictions of the reduced and full approaches will be visible whenever substrate concentrations are approximately equal to the Michaelis-Menten constant.


\newpage
\section*{\large \bf Supplementary information} 

\subsection{General formulation of master equations for elementary and non-elementary processes}

Consider a general chemical system confined in a compartment of volume $\Omega$ and consisting of a number $N$ of distinct chemical species
interacting via $R$ chemical reactions of the type
\begin{equation}
s_{1j} X_1 + \ldots + s_{Nj} X_{N} \xrightarrow{k_j} \ r_{1j} X_{1} + \ldots + r_{Nj} X_{N}. \label{eq1-supp}
\end{equation}
Here, $j$ is an index running from $1$ to $R$, $X_i$ denotes chemical species $i$, $s_{ij}$ and $r_{ij}$ are the stoichiometric coefficients, and $k_j$ is the macroscopic rate of reaction. Note that these reactions are not necessarily elementary (unimolecular or bimolecular reactions). If the $j$th reaction is elementary then its rate $k_j$ is a constant while if it is non-elementary $k_j$ is a function of macroscopic concentrations. The general form of the master equation for both cases is \cite{vanKampen-supp}
\begin{equation}
\frac{\partial P(\vec{n},t)}{\partial t} = \Omega \sum_{j=1}^{R} \biggl( \displaystyle\prod_{i=1}^N E_i^{-S_{ij}} - 1 \biggr) \hat{f}_j(\vec{n},\Omega) P(\vec{n},t), \label{eq2-supp}
\end{equation}
where $P(\vec{n},t)$ is the probability that the system is in a particular mesoscopic state $\vec{n}=(n_1,...,n_N)^T$ and $n_i$ is the number of molecules of the $i$th species. Note that $E_i^{x}$ is a step operator -- when it acts on some function of the absolute number of molecules, it gives back the same function but with $n_i$ replaced by $n_i + x$. The chemical reaction details are encapsulated in the stoichiometric matrix $S_{ij}=r_{ij}-s_{ij}$ and in the microscopic rate functions $\hat{f}_j(\vec{n},\Omega)$. The probability that the $j$th reaction occurs in the time interval $[t,t+dt)$ is given by $\Omega \hat{f}_j(\vec{n},\Omega) dt$.

For elementary reactions, the microscopic rate function takes one of four different forms, depending on the order of the $j$th reaction: (i) a zeroth-order reaction by which a species is input into a compartment gives $\hat{f}_j(\vec{n},\Omega)=k_j$; (ii) a first-order unimolecular reaction involving the decay of some species $h$ gives $\hat{f}_j(\vec{n},\Omega)=k_j n_h \Omega^{-1}$; (iii) a second-order bimolecular reaction between two molecules of the same species $h$ gives $\hat{f}_j(\vec{n},\Omega)= k_j n_h (n_h-1) \Omega^{-2}$; (iii) a second-order bimolecular reaction between two molecules of different species, $h$ and $v$, gives $\hat{f}_j(\vec{n},\Omega)=k_j n_h n_v \Omega^{-2}$. Note that these forms for the microscopic rate functions have been rigorously derived from microscopic physics \cite{Gillespie1992-supp, Gillespie2009-supp} and hence the validity of Eq.~(\ref{eq2-supp}) for elementary reactions is guaranteed.

For non-elementary reactions, the form of the microscopic rate function has to be basically guessed by analogy with the prescription for elementary reactions. For example, for the set of reactions (3) in the main text, the second reaction is a non-elementary first-order reaction with a time-dependent macroscopic rate constant $k'(t) = k_2 [E_T] / (K_M + [X_S(t)])$, where $[E_T]$ is the constant macroscopic total enzyme concentration and $[X_S(t)]$ is the instantaneous macroscopic concentration of species $S$. Hence, one would use the microscopic rate function $\hat{f} (\vec{n},\Omega)= k_2 [E_T] (n_S / \Omega) / (K_M + n_S / \Omega)$ based on the formula stated above for an elementary first-order reaction. Of course, master equations based on microscopic rate functions obtained from this procedure are ad-hoc and have no fundamental basis.

\subsection{General formulation of the linear noise approximation in
steady-state conditions}

Here, we provide a step by step recipe to construct the linear noise approximation (LNA) of the master equation, Eq.~(\ref{eq2-supp}), for the set of reactions (\ref{eq1-supp}). We note that this approximation is only valid for a monostable system [the condition is formally given by Eq.~(3.4) in Ch.~X of the book by van Kampen\cite{vanKampen-supp}]. Let the macroscopic steady-state concentration of species $i$ be given by $[X_i]$ and the derivative with respect to this variable be denoted by $\nabla_{i}$. Furthermore, we shall distinguish matrices by underlining them. The five steps to constructing the LNA in steady-state conditions (for both elementary and non-elementary reactions) are then as follows:

\begin{enumerate}
\item Construct the $N \times R$ stoichiometric matrix, $\matrix{S}$, whose
$i - j$ element is given by $r_{ij}-s_{ij}$.

\item Construct the macroscopic rate function vector $\vec{f}$ with elements
$f_j=k_j \prod_{m=1}^N ([X_m])^{s_{mj}}$ and the diagonal matrix
$\matrix{F}$ with elements $F_{ii} = f_i$.

\item Construct the Jacobian matrix $\matrix{J}$ whose $i - j$ element is
given by $\nabla_{j} (\matrix{S}.\vec{f})_i$. Construct the diffusion matrix
$\matrix{D}=\matrix{S} \cdot \matrix{F} \cdot\matrix{S}^T$.

\item The stochastic differential equations (linear Langevin equations)
approximating the chemical master equation for the set of reactions (\ref{eq1-supp}) in
the limit of large molecule numbers are given by \cite{ElfEhrenberg-supp}
\begin{equation}
\frac{\partial}{\partial t} \vec{\eta}(t) = \matrix{J} \cdot \vec{\eta}(t) + \Omega^{-1/2}
\matrix{S} \cdot \sqrt{\matrix{F}} \vec{\Gamma}(t), \label{eq3-supp}
\end{equation}
where $\eta_i(t)$, the $i$th entry of the vector $\vec{\eta}(t)$, denotes the fluctuations about the macroscopic steady-state concentration of species $i$, i.e., $\eta_i(t) = (n_i(t) / \Omega) - [X_i]$. The $R$ dimensional vector $\vec{\Gamma}(t)$ is white Gaussian noise defined by $\langle \Gamma_i(t) \rangle = 0$ and $\langle \Gamma_i (t)\Gamma_j (t') \rangle = \delta_{i,j} \delta(t - t')$.

\item The covariance matrix  $\matrix{\sigma}$ of the fluctuations in Eq.~(\ref{eq3-supp}) is obtained by solving the Lyapunov equation \cite{ElfEhrenberg-supp,Keizer-supp}
\begin{equation}
 \matrix{J} \cdot \matrix{\sigma}+\matrix{\sigma}\cdot \matrix{J}^T+\matrix{D}/\Omega=0,
 \end{equation}
where $\sigma_{ij}=\langle \eta_i \eta_j \rangle$. The variance of the fluctuations is hence given by the diagonal elements of $\matrix{\sigma}$.

 \end{enumerate}

\subsection{Parameter values used in stochastic simulations}

Parameter values for the stochastic simulations shown in Fig.~\ref{fig12}(a) are as follows. The enzymes, Malate dehydrogenase and Chymotrypsin, both have $\beta = 1$; we have simulated only the first of these enzymes by using $\Omega=17\,fL$, $[E_T]=10\,nM$, $k_2=5 \, s^{-1}$, $k_1=5\,s^{-1}$, and $k_0=5\times10^7 \, M^{-1}s^{-1}$. For the case $\beta = 2.8$ (the enzyme Lactate dehydrogenase), we used $\Omega=0.017\,fL$, $[E_T]=10\,\mu M$, $k_2=210 \, s^{-1}$, $k_1=75 \, s^{-1}$, and $k_0=3\times10^6 \, M^{-1}s^{-1}$. For the case $\beta = 10$ (a case compatible with a highly efficient enzyme), we used $\Omega=170 \, fL$, $[E_T]=1 \, nM$, $k_2=1 \, s^{-1}$, $k_1=0.1 \, s^{-1}$, and $k_0=10^8 \, M^{-1}s^{-1}$. In all cases the total number of enzyme molecules $n_T$ was $100$. Parameter values for Fig.~\ref{fig12}(b) are $\beta = 10$, $k_2=1 \, s^{-1}$, $k_1=0.1 \, s^{-1}$, $[E_T]=1 \, nM$, and $k_0=10^8 \, M^{-1}s^{-1}$.

The rate constants for the cases $\beta = 1$ and $\beta = 2.8$ were obtained from the experimental studies.\cite{Lodola1978-supp, Renard1973-supp, Boland1975-supp} The rate constants for $\beta = 10$ were not for a specific enzyme and hence were chosen from the known physiological ranges: for $k_2$ the range is\cite{Berg-supp} $1 - 10^4 \, s^{-1}$, for $k_0$ the range is\cite{Fersht-supp} $10^6 - 10^8 \, s^{-1} M^{-1}$, and for $K_M$ the range is\cite{Berg-supp} $10^{-1} - 10^{-7} \, M$. Similarly, the total enzyme concentrations were chosen from the physiological ranges: nano to millimolar concentrations.\cite{GrimaSchnell-supp} The compartment volumes for the data in Fig.~\ref{fig12}(a) were chosen such that the total number of enzyme molecules $n_T$ was $100$ in all cases; for Fig.~\ref{fig12}(b) the volumes were chosen such that $n_T$ could be varied over the range $1$ to $100$.

\end{document}